\documentclass[pdftex,twocolumn,showpacs,preprintnumbers,floatfix,prl]{revtex4-1}

\pdfoutput=1

\usepackage{amsmath,amssymb}
\usepackage{bm}
\usepackage{dcolumn}
\usepackage{graphicx}
\usepackage{hyperref}
\usepackage{rotating}
\usepackage{times}

\newcommand{\forH}{$^{14}$NH$_3$}
\newcommand{\mH}{NH$_3$}
\newcommand{\JK}{$\left|J,K\right>=\left|1,1\right>$}

\begin{document}
\title{Microwave lens for polar molecules}
\author{Hitoshi Odashima$^{1,2}$}
\author{Simon Merz$^1$}
\author{Katsunari Enomoto$^{1,3}$}
\author{Melanie Schnell$^{1,*}$}
\author{Gerard Meijer$^1$}
\affiliation{$^1$Fritz-Haber-Institut der Max-Planck-Gesellschaft,
  Faradayweg 4-6, D-14195 Berlin, Germany \\
$^2$Department of Physics, Meiji University, Kawasaki 214-8571, Japan \\
$^3$Department of Physics, University of Toyama, Toyama 930-8555, Japan}
\date{\today}

\begin{abstract}
We here report on the implementation of a microwave lens for neutral
polar molecules suitable to focus molecules both in
low-field-seeking and in high-field-seeking states. By using the
TE$_{11m}$ modes of a 12 cm long
cylindrically symmetric microwave resonator, Stark-decelerated
ammonia molecules are transversally confined. We investigate the
focusing properties of this microwave lens as a function of the
molecules' velocity, the detuning $\Delta$ of the microwave frequency
from the molecular resonance frequency, and the microwave
power. Such a microwave lens can be seen as a first important step
towards further microwave devices, such as decelerators and traps. 
\end{abstract}

\maketitle
Since the early 1920s, external electric and magnetic fields have
been used in ever more sophisticated setups to gain increasing
control over the molecules' degrees of freedom. As a first step, both an electric and
a magnetic deflector have been realized \cite{Wrede:ZP44:261,Stern:PZ23:476}, which allowed to select
single quantum states to perform
ground-breaking new experiments. In the mid fifties, the implementation of an electrostatic
quadrupole lens allowed Townes and coworkers to demonstrate the
MASER by exploiting the different focusing properties of the \JK~
inversion doublet levels of the vibronic ground
state of ammonia \cite{Gordon:PR95:282,Gordon:PR99:1264}. By now, it
has become possible to decelerate and
trap neutral molecules in the gas phase using switched electric or
magnetic fields, enabling novel experiments in the field of cold molecules
\cite{Meerakker:NatPhys4:595,Bell:MolPhys107:99,Schnell:ACIE48:6010}.

Besides external
electric and magnetic fields, also electromagnetic radiation, such as microwave and 
optical fields, can be used to manipulate the motion of molecules. For
example, an optical decelerator has been implemented \cite{Barker:PRA66:065402,Fulton:NatPhys2:465}. Optical
dipole traps are widespread in the field of cold atoms \cite{Grimm:AdvAtomMolOptPhys42:95}. 

In 1975, Hill and Gallagher implemented a deflector based on a
rectangular microwave resonator \cite{Hill:PRA12:451}, the
simplest molecule-manipulation device using microwave radiation. They detected the deflection of a
beam of CsF molecules, but their experimental results could not
be fully understood and they did not pursue these
experiments. 

In this Letter, we demonstrate and characterize a microwave lens for
polar molecules, which allows us to focus polar molecules using
microwave radiation. A packet of supersonically expanded ammonia
molecules (\forH) is decelerated to low velocities of around 20 m/s using a
Stark decelerator. Upon leaving the decelerator, they enter a
cylindrically symmetric microwave resonator. The applied microwave
field prevents the molecules from spreading out in
transverse directions, i.e., it acts like a positive lens on polar
molecules.  For this, we exploit
the same inversion transition as Townes and coworkers used in their
original MASER experiments.

The force
exerted
on a molecule in microwave fields depends on its 
dipole moment, the sign and the magnitude of the detuning of the microwave 
frequency from the molecular resonance frequency, and the
microwave field strength. High input powers and
high-Q cavities are available, so that, in principle, large field
strengths can be achieved. Furthermore, in contrast to optical fields, very small detunings can
be exploited due to the long lifetime of the states coupled in the
microwave transition. Such small
detunings in turn lead to strong interaction forces. 
In addition, for microwave manipulation the direction
of the interaction force (AC Stark effect \cite{Autler:PR100:703}) depends on the sign of the
detuning. By choosing the corresponding detuning, both 
molecules in low-field-seeking and in high-field-seeking states can 
be affected.  
This is particularly advantageous for large molecules and
molecules in their internal ground state, which are always
high-field seeking in DC electric fields. To focus, decelerate and
trap them using electric fields, dynamic focusing has to be applied \cite{Bethlem:PRL88:133003,Veldhoven:PRL94:083001},
which leads to stringent experimental demands. Here, microwave fields
could be a promising alternative. The microwave lens
demonstrated in this work can be seen as an important step towards the
realization of microwave deceleration and trapping devices, which
have been suggested recently \cite{DeMille:EPJD31:375,Enomoto:PRA72:061403}. 

\begin{figure}
   \begin{center}
   \includegraphics[width=0.9\linewidth]{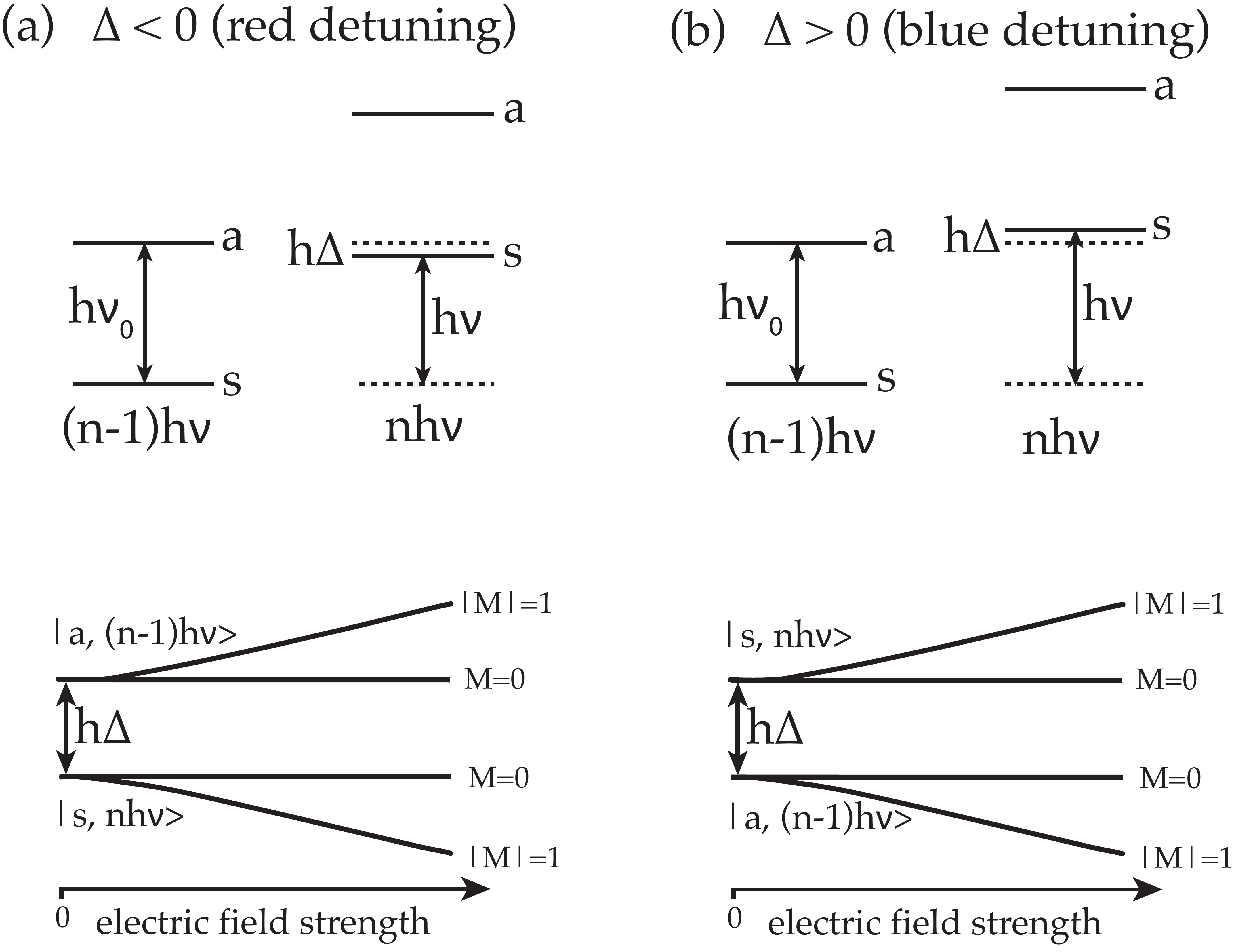} 
   \caption{Dressed-state picture and AC Stark shift
     of the \JK~ inversion doublet of
     \forH~for (a) red ($\Delta<0$) and (b) 
     blue ($\Delta>0$) detuning of the microwave frequency $\nu$ relative to the
 molecular resonance frequency $\nu_0=23695$ MHz. Depending on the
detuning, either the symmetric (\textit{s}) lower level ($\Delta<0$) or the
antisymmetric (\textit{a}) upper level ($\Delta>0$) of the inversion
doublet becomes high-field seeking in microwave fields.}
   \label{fig:Stark}
   \end{center}
\end{figure}

The interaction energy of polar molecules with microwave radiation is
based on the AC Stark effect. For ammonia, the energy of the inversion
doublet levels as a function of the microwave field strength (Figure \ref{fig:Stark}) can be
written to a good approximation as:
\begin{equation}
	\label{equation:AC}
   E_{AC}(\rho,\varphi,z)=\frac{h\Delta}{2}\pm
   \sqrt{\left(\frac{h\Delta}{2}\right)^2+\left(\frac{E_0(\rho,\varphi,z) \mu MK}{2J(J+1)}\right)^2},
\end{equation}
with the molecular dipole moment $\mu$, the
total angular momentum quantum number $J$ and the projections of the
total angular momentum onto the molecular symmetry axis, $K$, and the
electric field axis, $M$, respectively. $\Delta = \nu-\nu_0$ is the so-called detuning defined as the
difference of the applied microwave frequency $\nu$ from the molecular
resonance frequency $\nu_0\approx23695$ MHz (inversion transition of
the \JK~level of the
vibronic ground state of \forH) and $h$ is Planck's
constant. $E_0(\rho,\varphi,z)$ is the electric field strength of the
microwave field, as defined below. Its maximum value
$\epsilon_{MW}$ is given by $\epsilon_{MW}\approx \sqrt{Q_0U_0^2
  P/(\pi\nu \epsilon_0)}$ \cite{Grabow:RSI76:093106}.  
Here, $Q_0$ is the quality factor of the resonator, and $U_0^2$ is a geometrical factor which can be approximated by
$4.2/V$ for TE$_{11m}$ modes (\textit{vide infra}) in a cylindrically symmetric resonator, with
$V$ being the resonator volume. $P$ is
the microwave power coupled into the resonator and $\epsilon_0$ the permittivity of free
space.
Consequently, the focusing force in a microwave lens depends on the molecular dipole
moment $\mu$, the resonator properties $Q_0$ and $U_0$, the microwave
power $P$, and
the sign and the magnitude of the detuning $\Delta$.

While the magnitude of $\Delta$ determines the strength, its sign defines the direction of the
focusing force, i.e., it determines whether a particular molecular state is
attracted by a field maximum or a field minimum (eq. \ref{equation:AC}). This is schematically displayed in Fig.~\ref{fig:Stark} for
the antisymmetric (\textit{a}) upper (low-field seeking in DC
electric fields) and the symmetric (\textit{s}) lower energy levels (high-field
seeking in DC electric fields) of the \JK~inversion doublet of \forH~molecules (see, for
example, ref. \cite{Bethlem:PRA65:053416}). In the dressed-state
picture, the molecular energy levels are ''dressed'' with the number
of photons $n$. This way, the \textit{s} state dressed with $n$ photons can
come close to the \textit{a} state with $(n-1)$ photons, resulting in a strong
interaction. For small detunings, this leads to a linearization of
the AC Stark energy. While for red detuning ($\Delta
< 0$) the \textit{a} state is low-field seeking, it becomes high-field
seeking for blue detuning ($\Delta > 0$).
Consequently, by simply changing the sign of
the detuning $\Delta$ a particular state can be either high-field or low-field seeking in microwave fields. This property
can be exploited for various applications, such as the manipulation of
larger and more complex molecules. Here, we
Stark decelerate \forH~molecules in the \textit{a} state, which is
low-field seeking in DC electric fields. Subsequently, we can conveniently focus these molecules
using blue detuned microwave radiation without having to transfer the
molecules to a different state. Although such transfer can be performed in
principle, it is accompanied by
significant losses, as the population is distributed over a large number
of hyperfine levels \cite{Veldhoven:PRL94:083001}.

\begin{figure}
   \begin{center}
   \includegraphics[width=0.9\linewidth]{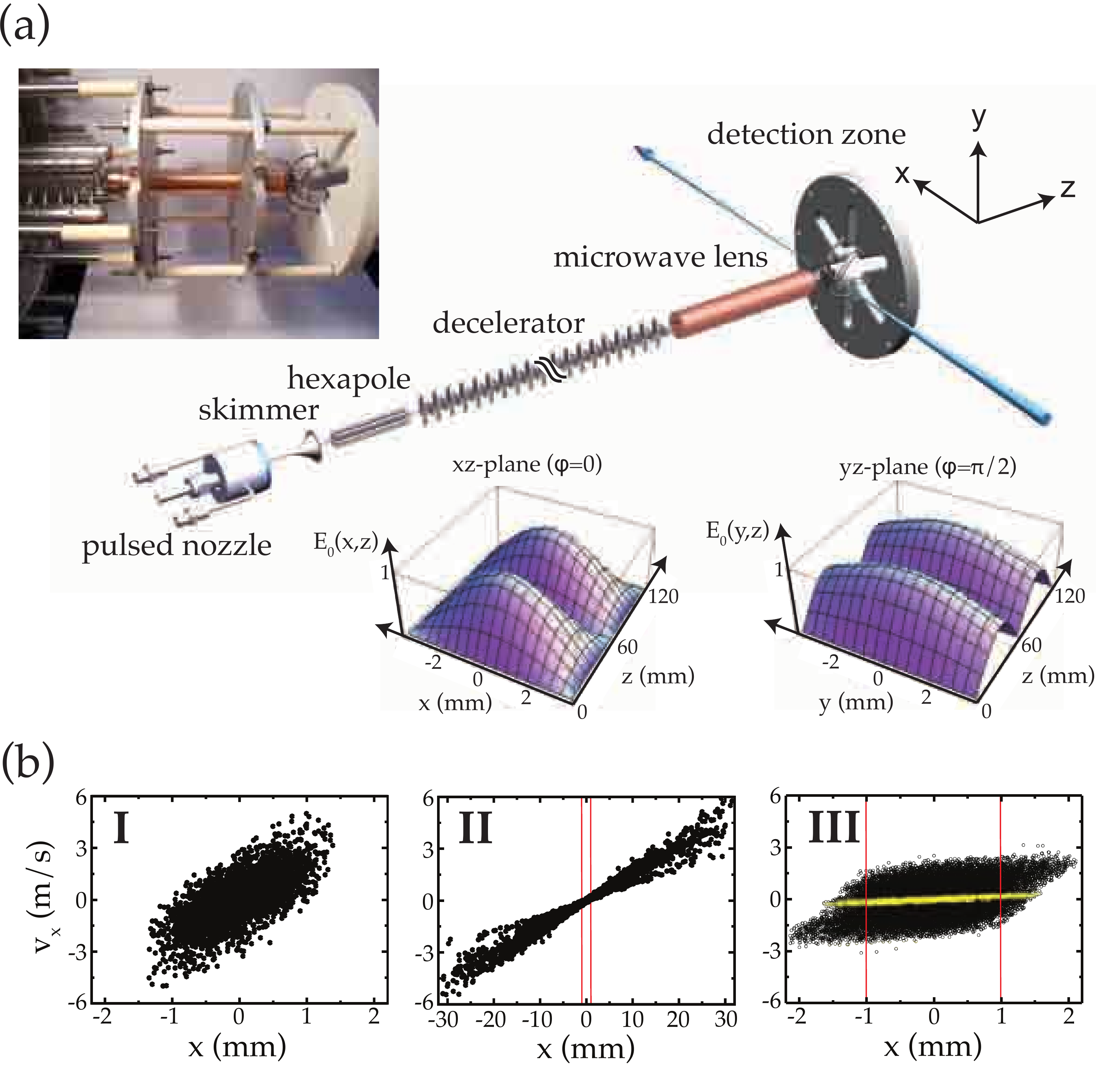} 
   \caption{(a) Scheme of the experimental setup, complemented by a
     photo showing the end of the decelerator, the microwave resonator
     and the detection zone. The calculated electric field strength $E_0(\rho,\varphi,z)$ for the TE$_{112}$ mode is shown
     for two perpendicular planes of the resonator ($E_0(x,z)$ for $\varphi=0$ and $E_0(y,z)$ for
     $\varphi=\pi/2$). Here, $z$ is the symmetry axis of the resonator
     which coincides with the molecular beam axis. Note the different
     scales for the transverse and longitudinal directions.
(b) Simulated transverse phase-space
     distributions of the molecules in the decelerated packet (i.e.,
     their positions $x$ and velocities v$_x$): (I) at
     the beginning of the microwave resonator, (II)  in the laser
     detection region if no microwave resonator
     would be integrated, and (III) at the same position but including
     the microwave resonator, with microwave radiation turned on
     (black circles, $P=3.5$ W) and turned off (yellow circles). In (II) and (III), 
     the detection region along $x$ is also indicated by vertical red
     lines. Note the
     different scales of the $x$ axes for (II) and (III).}
   \label{fig:exp}
   \end{center}
\end{figure}

The experimental setup is schematically shown in Fig.~\ref{fig:exp}(a). 
A detailed description of the molecular beam machine and in particular of the 
deceleration of a beam of ammonia molecules is given elsewhere
\cite{Bethlem:PRA65:053416}. The 120 mm long microwave resonator is
made of copper and its entrance is located about 2.5 mm behind the end of the
decelerator. It has an inner radius of 3.705 mm. The resonator is closed by two endcaps with 3 mm holes
to allow the molecules to fly through. The microwave radiation is coupled
into the resonator using a dipole
antenna of 5 mm length and 0.4 mm diameter, which is oriented along
the y direction. 
The electric field
component of the microwave field propagates parallel to the antenna, i.e.,
the electric field is linearly polarized perpendicular to the resonator
axis (TE mode). 
In the experiment, we use TE$_{11m}$ modes for focusing ammonia
molecules. Here, 1,1, $m$ denote the number of maxima in radial
($\rho$), azimuthal ($\varphi$), and longitudinal (z) direction, respectively.
Since the antenna is located at half the length of the resonator,
TE$_{11m}$ modes with odd $m$, which have an electric field maximum at
this position, are strongly perturbed.  Here, we use the TE$_{112}$
mode ($\Delta$=+71 MHz) and the TE$_{114}$ mode ($\Delta$=+459 MHz) to
confine the transverse motion of ammonia molecules. For these modes,
$Q_0$ is about 5000 \cite{Lamont:Waveguides:3}. 

The distribution of the electric field strength (mode pattern) for TE$_{11m}$ modes in cylindrical coordinates
can be calculated using
\begin{equation}
	\label{equation:E_feld2}
 \begin{split}
  E_{MW}(\rho,\varphi,z,t)=E_0(\rho,\varphi,z)\cos(\omega t+\phi) \\
= \epsilon_{MW}
  \frac{U(\rho,\varphi,z)}{U_0}\cos(\omega t +\phi),
 \end{split}
\end{equation}
with $U(\rho,\varphi,z)$ giving the field distribution \cite{Jackson:1998}:
\begin{equation}
 \begin{split}
U(\rho,\varphi,z)= 2U_0 \sqrt{\sin^2{\left(\frac{m\pi z}{d}\right)}} \;
\times \\\sqrt{\left[\frac{J_1(\gamma\rho)}{\gamma\rho}\right]^2\sin^2{\varphi} + \left[ \frac{\partial J_1(\gamma\rho)}{\gamma\partial
      \rho}\right]^2\cos^2{\varphi}} \; .
 \end{split}
\label{equation:E_feld3}
\end{equation}
Here, $J_1(\gamma\rho)$ is the Bessel function of first order, $d=120$
mm is the length of the resonator.
$\gamma$ is defined as $\gamma=x'_{11}/R$,
with $x'_{11}$=1.841 being the first root of the derivative of $J_1(\gamma\rho)$ (for TE$_{11m}$
modes) and $R$=3.705 mm being the radius of the resonator.

In the inset of Fig.~\ref{fig:exp}(a), the calculated electric field strength
$E_0(\rho,\varphi,z)$  is shown for the TE$_{112}$ mode for two perpendicular planes; $E_0(x,z)$ for $\varphi=0$ 
and $E_0(y,z)$ for $\varphi=\pi/2$. 
The electric field distribution of the TE$_{112}$ mode exhibits an
electric field maximum on the resonator axis z, which decreases
towards the resonator walls in transverse directions. Consequently, molecules in AC
high-field-seeking states are transversally focused towards the
resonator axis $z$. Note that the electric field distribution is
not cylindrically symmetric, which leads to different gradients and
thus different focusing strengths for the two transverse
directions $x$ and $y$.

In the experiment, decelerated packets of \mH~molecules leave the decelerator 
in the \textit{a} state of the \JK~inversion doublet
with a mean longitudinal velocity $\bar{v}_z$ of, typically, 
20 m/s, and with a full width at half maximum (FWHM) velocity spread of about
10~m/s. The transverse velocity distribution (FWHM) is about 5
m/s. The decelerated packet leaving the decelerator contains
approximately 10$^6$ molecules and has a full spatial extent of about
2 mm along all directions. 

While proceeding to the entrance of
the microwave resonator in free flight, the decelerated
packet spreads out in all directions (Fig.~\ref{fig:exp}(b), I). If no microwave resonator would be integrated into the setup, the
molecules would have to proceed another 128.5 mm of free flight to the
detection region, where they are state-selectively detected using a laser-based ionization
detection scheme. The laser beam with a waist of about 100 $\mu$m is
oriented parallel to the x axis. In the x direction, 
the ion extraction setup limits the detection region to about 2
mm. During this free flight, the packet of molecules spreads out
dramatically (Fig.~\ref{fig:exp}(b), II), so that only a small
fraction of molecules would still be detected. However, for the setup with microwave resonator, the molecular cloud is 'skimmed' by the
entrance and exit holes of the resonator (yellow circles in
Fig.~\ref{fig:exp}(b), III): only the molecules with very low transverse velocities are able to pass. If the
microwave field is turned on with $P=3.5$ W coupled into the resonator, the packet of slow ammonia
molecules is kept together in transverse directions
(Fig.~\ref{fig:exp}(b), III, black circles). The maximum accepted transverse velocity
of our microwave lens for 3.5 W is about $\pm3$ m/s in $x$ direction and
about $\pm1.6$ m/s in $y$ direction. The microwave field is
turned on as long as the decelerated
packet is in the resonator, i.e., for focusing molecules
with $\bar{v}_z$=20 m/s, the microwave field is on for 6 ms. Note that in the
deceleration process of ammonia as applied here, three decelerated
packets which are trailing each other by 11 mm are emitted from the
Stark decelerator \cite{Heiner:PCCP8:2666}. While flying through the
microwave resonator, these three packets will merge.

\begin{figure}
   \begin{center}
   \includegraphics[width=0.85\linewidth]{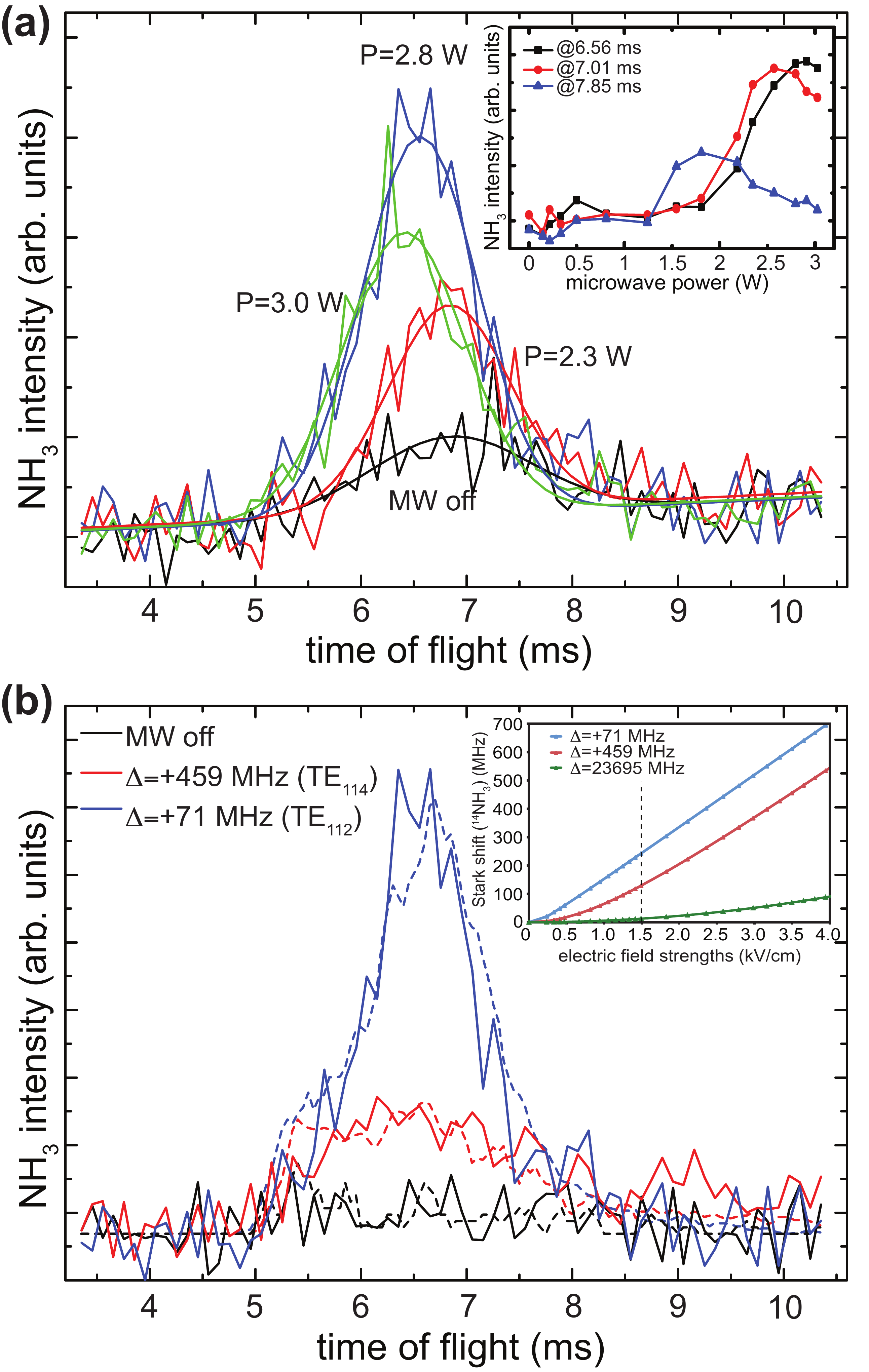} 
   \caption{Experimental results for focusing ammonia molecules
     (\forH) in
     their \JK~state decelerated to $\bar{v}_z$=20 m/s. (a) Time-of-flight measurements along with
     Gaussian fits for
     different microwave powers $P$ using the TE$_{112}$ mode
     ($\Delta=+71$ MHz). The inset shows the \forH~intensity as a function of $P$ for different detection times. (b) Time-of-flight measurements (solid curves) for two different detunings ($\Delta=+71$ MHz and +459 MHz, respectively)
   using $P=2.8$ W and 0 W. The dashed curves correspond to Monte-Carlo simulations. The inset displays the AC Stark
     shift (absolute values) for \forH~for two different detunings
     $\Delta$ calculated using
     eq. \ref{equation:AC} along with the corresponding DC Stark
     shift.}
   \label{fig:measurement}
   \end{center}
\end{figure}

Figure \ref{fig:measurement} shows the experimental results for
focusing ammonia molecules in their \JK~state. Fig. \ref{fig:measurement}(a) displays time-of-flight
measurements for \forH~molecules decelerated to $\bar{v}_z$=20 m/s for different
microwave powers $P$, along with
Gaussian fits to the observed signal curves. At $t=0$, the decelerated
packet leaves the decelerator, i.e., the high voltages at the
decelerator are turned off. The signal intensity is
increased by about a factor of 8 due to microwave focusing.
Furthermore, for increasing power, higher velocities are focused best: compared to the
measurements for 2.3 W ($\epsilon_{MW}$=1.33 kV/cm), the maxima of the
time-of-flight measurements are shifted to earlier arrival times, and
thus higher $\bar{v}_z$ values, by 0.28 ms for 2.8 W
($\epsilon_{MW}$=1.47 kV/cm) and by 0.43 ms for 3.0 W
($\epsilon_{MW}$=1.52 kV/cm). The highest power of 3.0 W causes
overfocusing for the most abundant velocity in the initial velocity
distribution, resulting in a reduced signal intensity compared to
2.8 W.

 Both the shape of the curves as well as the
position of the maxima are reproduced well by Monte Carlo trajectory
simulations (see also Figure \ref{fig:measurement}(b), dashed
curves). However, the intensity increase due
to microwave focusing is predicted to be about one order of
magnitude larger than experimentally observed. This is a strong
indication for significant losses in the experiment. One major loss
channel might be nonadiabatic transitions \cite{Kirste:PRA79:051401}
to $M=0$ states in the
low-field regions close to the electric field nodes at the beginning, center
and end of the resonator. One possible way to reduce these losses would
be to use the TE$_{111}$ mode with only one maximum along the
longitudinal axis. 

The inset of Fig.~\ref{fig:measurement}(a) shows the dependence of the
observed signal on the microwave power $P$. The three
measurements shown have been performed for different detection
times (6.56 ms, 7.01 ms, and 7.85 ms). 
For all curves, a steep increase in signal
intensity is found with increasing microwave power. For the
measurements taken at 7.01 ms, a clear maximum is obtained at around
2.5 W. For higher microwave power, overfocusing is observed. For a later
detection time (7.85 ms) corresponding to slower molecules,
the maximum is clearly shifted to lower power (1.8 W), while for an
earlier detection time (6.56 ms) and thus faster molecules, it is
shifted to higher $P$ values (around 3 W).

In Fig.~\ref{fig:measurement}(b), time-of-flight measurements for $\bar{v}_z$=20
m/s slow \forH~molecules using a
constant microwave power of 2.8 W ($\epsilon_{MW}$=1.47 kV/cm), but two different detunings are
shown. The solid curves are measurements, while the dashed curves
correspond to trajectory simulations, which agree well with the
experimental results. In the experiment, the signal
increase due to focusing amounts to only about a factor of 2 for the
TE$_{114}$ mode
compared to a factor of 8 for the TE$_{112}$ mode. This can be explained
with a higher probability of nonadiabatic transitions for the
TE$_{114}$ mode due to the larger number of low-field regions, and the
smaller focusing force for $\Delta$=+459 MHz compared to
$\Delta$=+71 MHz due to the smaller slope of the Stark shift; 141
MHz/(kV/cm) and 182 MHz/(kV/cm) at 1.5 kV/cm, respectively  (inset of
Fig. \ref{fig:measurement}(b)). Note that for the corresponding DC Stark effect the
slope of the Stark energy at 1.5 kV/cm is one order of magnitude
smaller (17 MHz/(kV/cm)).

In summary, we have focused ammonia molecules using a
novel microwave lens. The focusing force depends on the particular
state, the molecular dipole moment, the sign and magnitude of the
detuning of the applied microwave frequency from the molecular
resonance frequency, and the microwave
field strength. Such a microwave lens can be seen as a first important step towards further
manipulation devices based on microwave radiation, such as a
microwave decelerator and trap. Microwave manipulation is a promising
route to gain complete control over larger and more complex
molecules.

\begin{acknowledgments}
The authors thank Jens-Uwe
Grabow, John M. Doyle, Jochen K\"upper and Boris G. Sartakov for valuable scientific
discussions and Henrik Haak for
technical support. H.O. thanks Meiji University for granting him a
sabbatical leave at FHI. M.S. acknowledges a Liebig 
grant from the \textit{Fonds der Chemischen Industrie}. 
\end{acknowledgments}

$^*$corresponding author: schnell@fhi-berlin.mpg.de

\bibliography{string,mp}
\bibliographystyle{jk-apsrev}
\end{document}